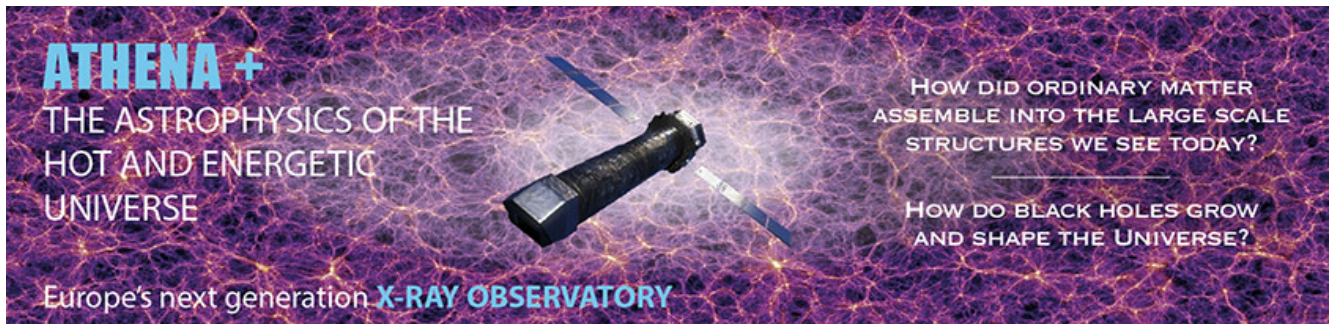

# The Hot and Energetic Universe

An *Athena+* supporting paper

# The formation and growth of the earliest supermassive black holes

## Authors and contributors

**James Aird, Andrea Comastri**, Marcella Brusa, Nico Cappelluti, Alberto Moretti, Eros Vanzella, Marta Volonteri, David Alexander, Jose Manuel Afonso, Fabrizio Fiore, Ioannis Georgantopoulos, Kazushi Iwasawa, Andrea Merloni, Kirpal Nandra, Ruben Salvaterra, Mara Salvato, Paola Severgnini, Kevin Schawinski, Francesco Shankar, Cristian Vignali, Fabio Vito

# 1. EXECUTIVE SUMMARY

A crucial challenge in astrophysics over the coming decades will be to understand the origins of supermassive black holes (SMBHs: $M_{BH} \sim 10^{6-10} M_\odot$) that lie at the centres of most, if not all, galaxies. The processes responsible for the initial formation of these SMBHs and their early growth via accretion – when they are seen as Active Galactic Nuclei (AGNs) – remain unknown. The *Athena+* next-generation X-ray observatory will be uniquely capable of tracing the accretion activity of the earliest SMBHs at $z>6$ and will address the following science questions:

- How do the "seeds" of the first SMBHs form?
- What processes drive the initial growth of SMBHs and trigger AGN activity in the early Universe?
- What influence do early AGNs have on structure formation, reionisation of the Universe and the evolution of the first galaxies?

To answer these questions it is necessary to identify large populations of AGNs at redshifts $z>6$, probing the epoch where the first galaxies and SMBHs formed. The known population of AGNs at $z>6$ currently consists of a few tens of luminous ($L_{bol}>10^{47}$ erg s$^{-1}$) optical quasars identified using large optical or near-infrared imaging surveys (e.g. Fan et al. 2003), with the highest redshift source at $z = 7.085$ (Mortlock et al. 2011). However, to fully reveal the growth mechanisms of early SMBHs, **we must identify lower luminosity and obscured $z>6$ AGNs, which represent the bulk of early SMBH growth**. Sensitive X-ray observations are required to identify these early AGNs and accurately trace their accretion power. Hard X-ray light can penetrate the large columns of gas and dust obscuring the growing SMBH. Furthermore, it is a unique signpost of accretion activity, uncontaminated by star formation processes, which prevent reliable AGN identification at other wavelengths (e.g. optical, infrared).

*Athena+* will enable X-ray surveys to be carried out **two orders of magnitude faster** than with *Chandra* or *XMM-Newton* due to the combination of a large (~2m²) collecting area at ~1keV, a large field-of-view (40'x40'), and a sharp (<5") PSF over a substantial fraction of the field-of-view. This capability allows *Athena+* to access a new discovery space. A multi-tiered *Athena+* Wide Field Imager (WFI) survey, combining deep and moderate depth exposures for a total observing time of 25Ms (i.e. 1 year, assuming an 80% observing efficiency), is expected to identify over 400 $z=6-8$ X-ray selected AGNs and around 30 at $z>8$. Detection of low-luminosity AGNs at such high redshifts will provide direct constraints on the minimum mass of SMBH seeds and constrain their growth mechanisms. These X-ray surveys will pinpoint active SMBHs (to ~1" positional accuracy) within samples of $z>6$ galaxies identified by the large, state-of-the-art optical and near-infrared imaging surveys (from LSST, Euclid and Hyper-SuprimeCam) that will be available in the late 2020s. Further follow-up of these X-ray AGNs with E-ELT, ALMA and JWST will confirm their redshifts and yield stellar masses, star formation rates, cold gas properties, dust masses, and other important properties of the host galaxies. We will thus be able to determine the physical conditions within the host galaxies of early SMBHs, which is vital for understanding how SMBHs form, what fuels their subsequent growth, and to assess their impact on the early Universe. Follow-up of samples of $z>6$ galaxies with the *Athena+* X-ray Integral Field Unit (XIFU) could also reveal the presence of highly obscured AGNs, thanks to the detection of strong iron lines. Thus, *Athena+* will enable the first quantitative measurements of the extent and distribution of SMBH accretion in the early Universe.

# 2. THE INITIAL GROWTH OF SUPERMASSIVE BLACK HOLES

While much progress has been made on characterising and understanding the growth of SMBHs, with recent efforts focussing at the peak of both galaxy and SMBH growth at $z=1-4$ [1], how SMBHs form and grow in the earliest galaxies at $z>6$ is currently unknown. Two possible formation mechanisms have been proposed: 1) the seeds of early SMBHs are the remnants of the first generation of massive, low-metallicity stars (Population III stars) whose evolutionary end-point is thought to be a black hole with a mass of $\sim 10-100\ M_\odot$; 2) early SMBHs form from the direct collapse of primordial gas clouds, producing seeds with masses of $\sim 10^4-10^6\ M_\odot$. After forming the SMBH seeds can grow to higher masses either by merging or through accretion. Indeed, a sequence of frequent periods of intense accretion triggered by galaxy mergers is required to build up the large masses ($\sim 10^9 M_\odot$) of currently known $z>6$ SMBHs (e.g. Li et al. 2007), which are found powering extremely luminous quasars out to $z\sim 7$ (e.g. Fan et al. 2003, Mortlock et al. 2011). Large-scale inflows of cold gas, which assemble the first galaxies, may also fuel ongoing, high rates of accretion for these early SMBHs (Di Matteo et al. 2012; Dubois et al. 2013). Different seeding mechanisms require different levels of subsequent growth: Population III seeds must grow rapidly shortly after they have formed in early galaxies at $z$~20–30, whereas the more massive seeds from monolithic collapse can grow through extended periods of more moderate accretion. Understanding this early growth remains a major challenge in astrophysics.

---

[1]Note, with limited success for heavily obscured AGNs (see Georgakakis, Carrera, et al. 2013, *Athena+* supporting paper)

To constrain models of early SMBH formation and growth we must identify AGNs at $z>6$ and accurately measure the distribution of their accretion power (i.e. their luminosity function). Searches for luminous optical quasars at $z>6$ have placed important constraints on models of the early growth of SMBHs. However, the masses of the SMBHs powering these luminous quasars are several orders of magnitude larger than could be produced by any known seed mechanism. In fact, such quasars are identified based on their high luminosities, which require in turn masses in excess of $10^9$ $M_\odot$. Any dependence on the initial conditions is erased as the final mass is effectively determined by their sustained accretion, or through several mergers, while the initial seed mass and its early growth represent a tiny fraction of the final mass. To accurately track the growth of early SMBHs – and thus provide meaningful constraints for models of seed formation and how most SMBHs increase their masses – we must probe more typical, lower mass SMBHs. **Thus, we must identify lower luminosity and obscured $z>6$ AGNs, which correspond to the majority of early SMBH growth.**[2]

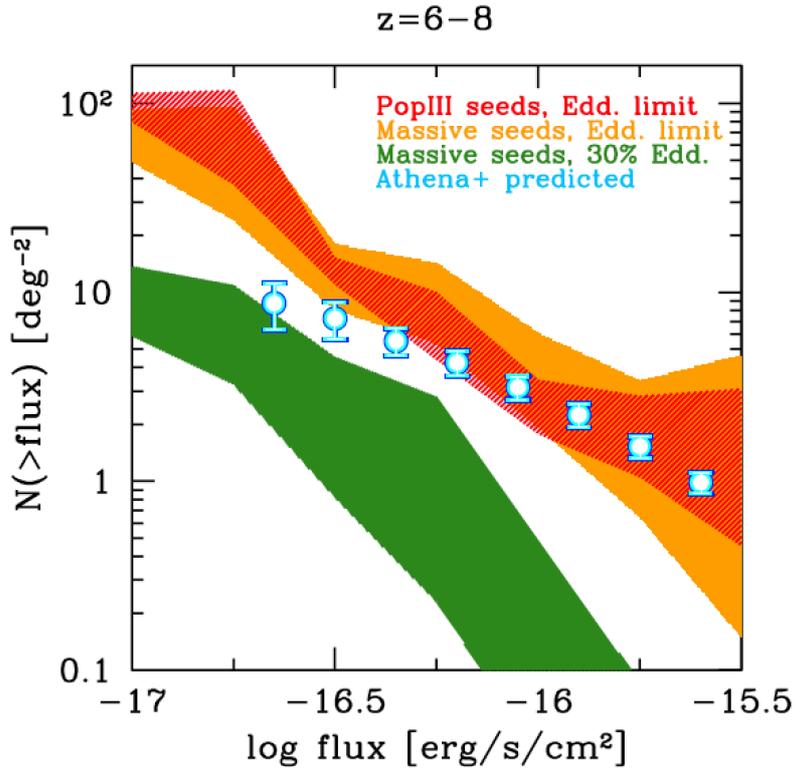

**Figure 1: Expectations for the number counts of high-redshift AGNs ($z$=6–8).** The blue circles show predicted measurements with *Athena+*, based on extrapolations from lower z; the error bars are based on the predicted number of detected sources with the 1-year *Athena+* survey described in Section 4. The shaded regions show predictions based on theoretical models (see Volonteri & Begelman 2010) that differ by black hole formation mechanism and growth rate (red: Population III star remnants, Eddington accretion; orange: massive seeds, Eddington accretion; green: massive seeds, 30% Eddington accretion). All of the models are built to produce billion solar masses SMBHs by $z$=6 (which are found powering luminous QSOs) and to be in satisfactory agreement with the luminosity function of AGNs at $z$=0.5–6 as well as the SMBH mass density at $z<3$ and the M-σ relation at $z$=0. The sizes of the regions correspond to current uncertainties (for a given seed/growth prescription) related to the limited volume that can be currently modeled and will be reduced by future simulations that combine high resolution with large volumes. The figure indicates the constraints that *Athena+* will provide on different physical models for early SMBH growth and formation – for example, finding a low number density of $z$=6–8 AGNs would imply low growth rates and rule out PopIII seeds, which must grow at high accretion rates to satisfy current observational constraints.

---

[2] We also note that recent observations (Schawinski et al. 2011) and theoretical considerations (Volonteri & Begelmann 2010) suggest that some seed black hole formation might be delayed until low redshift ($z$~1). *Athena+* will allow us to identify very low mass (~$10^4$-$10^6$ $M_\odot$) growing black holes in star-forming galaxies at $z$~1–3 enabling us to characterise their birth environments and thus constrain formation mechanisms.

In addition, feedback from AGNs may be playing a crucial role in shaping the evolution of galaxies. The impact of AGN feedback on the $z>6$ Universe and whether SMBHs and galaxies are co-evolving at these early times is completely unknown. To make progress requires studies of the extent and distribution of AGN activity in $z>6$ galaxies and the relationship to key galaxy properties such as stellar or gas mass, and star-formation rate. Furthermore, the $z\sim6$–10 range corresponds to the later stages of the epoch of reionisation (Dunkley et al. 2009). While the radiation from early stars dominates the re-ionisation process (e.g. Meiksin 2005), AGNs may play a significant role in these final stages of re-ionisation, can lead to more extended re-ionisation (due to the greater penetrating power of X-rays), and may help maintain re-ionisation at later times (e.g. Volonteri & Gnedin 2009, Faucher-Giguère et al. 2008). To constrain the role of AGNs in re-ionisation it is essential to measure the extent of AGN activity, traced by their luminosity function, during this crucial transition from the cosmic dark ages to the present Universe.

To directly identify large, statistical samples of typical luminosity $z>6$ AGNs, track their growth, constrain models of their fuelling mechanisms, and assess their impact on the early Universe requires a next-generation X-ray observatory that can perform surveys to comparable depths to the deepest *Chandra* fields, but over substantially larger areas. The *Athena+* X-ray observatory has the capability to probe this new discovery space and revolutionise our understanding of early SMBHs (see Figure 1 and Section 4 below).

## 3. IDENTIFYING HIGH REDSHIFT ACTIVE GALACTIC NUCLEI

To perform a census of AGN activity in the early Universe requires a selection technique that traces the bolometric luminosity of the system, while minimising the effects of obscuration and contamination from star formation. Many techniques have been developed over the last decade or so to identify AGNs over a wide range of redshifts and track their accretion power (e.g. rest-frame UV-optical colour, radio, mid-IR continuum or emission lines such as [OIV] $\lambda 25.89\mu m$ or [NeV] $\lambda 14.32\mu m$, or narrow optical emission lines such as [OIII] $\lambda 5007\text{Å}$). However, each of them turns out to be effective at identifying AGNs over a narrow range of the parameter space depending on the level of host star formation, host versus AGN luminosity, emission line intensities, IR and/or bolometric AGN luminosities and obscuration properties. None of them is able to provide a full census of the AGN population. Thus, while future facilities operating at optical, infrared and radio wavelengths (e.g. E-ELT, Euclid, JWST, SPICA, SKA) can employ these techniques to identify AGNs, they will be severely limited by the biases described above. Furthermore, accessing the necessary rest-frame wavelengths at $z>6$ requires observations at much longer wavelengths, in many cases limiting the sensitivity that can be achieved, even with the state-of-the-art facilities that will be available through the 2020s.

X-ray observations, on the other hand, have proven uniquely capable of efficiently identifying AGNs, determining their accretion power, measuring their luminosity function and thus tracing the accretion history of the Universe (e.g. Ueda et al. 2003, Aird et al. 2010). Hard X-ray light (shifted to lower, more easily observed energies when observing the highest redshift objects) can penetrate obscuring material that often accompanies lower luminosity AGNs. Furthermore, the hard X-ray emission is dominated by the AGN, providing the most reliable tracer of the bolometric luminosity, which lacks the strong contamination from the host galaxy light that plagues AGN identification at other wavelengths. Thus, X-rays are able to efficiently reveal the presence of an AGN and measure its luminosity, with longer wavelength facilities only required to provide redshifts and host galaxy properties (see Section 5 below). However, current deep and large area X-ray surveys with *Chandra* or *XMM-Newton* (e.g. Xue et al. 2011, Civano et al. 2011, Trichas et al. 2012) have yielded only a handful of directly detected AGNs at $z=5$–6 (predominantly with $L_X > 10^{44}$ erg s$^{-1}$) and no confirmed X-ray selected object at $z>6$. Thus, the levels of early SMBH growth remain unclear and poorly constrained – *Athena+* is required to reveal these early AGNs and provide the transformational leap in our understanding of the physics of SMBH formation and growth in the high-redshift Universe.

## 4. PROSPECTS FOR ATHENA+

Here we outline the prospects for a multi-tiered *Athena+* WFI survey, combining extremely deep and shallower wide-area WFI observations, that will identify AGNs at very high redshifts ($z>6$), measure their X-ray luminosity function (XLF), and thus revolutionise our understanding of the $z>6$ Universe. To assess the discovery space that must be probed, we require predictions of the evolution of the XLF at very high redshifts. Recent studies of AGNs at $z \sim 4$-5 with the latest *XMM-Newton* and *Chandra* surveys (Brusa et al. 2009, Civano et al. 2011, Fiore et al. 2012) have found there is a rapid drop in the space densities of both moderate and luminous AGNs above $z\sim3$. We model this behaviour and extrapolate it to higher redshifts by adjusting the Luminosity And Density Evolution (LADE) parameterisation of the evolution of the XLF at lower redshifts ($z<3$) from Aird et al. (2010). In Figure 2 we use this model to calculate the area that must be covered to a given flux limit in order to detect >10 high-$z$ AGNs in the indicated redshift ranges. A survey that tracks these curves will uniformly populate the XLF with ~ 10 sources per bin at the given redshifts. To constrain the faint end of the XLF (~$10^{43-44}$ erg s$^{-1}$) requires a survey that reaches flux limits of ~$3 \times 10^{-17}$ erg s$^{-1}$ cm$^{-2}$ over an area a few square degrees and flux limits of ~$2 \times 10^{-16}$ erg s$^{-1}$ cm$^{-2}$ over tens of square degrees. This coverage is beyond the capability of current X-ray observatories.

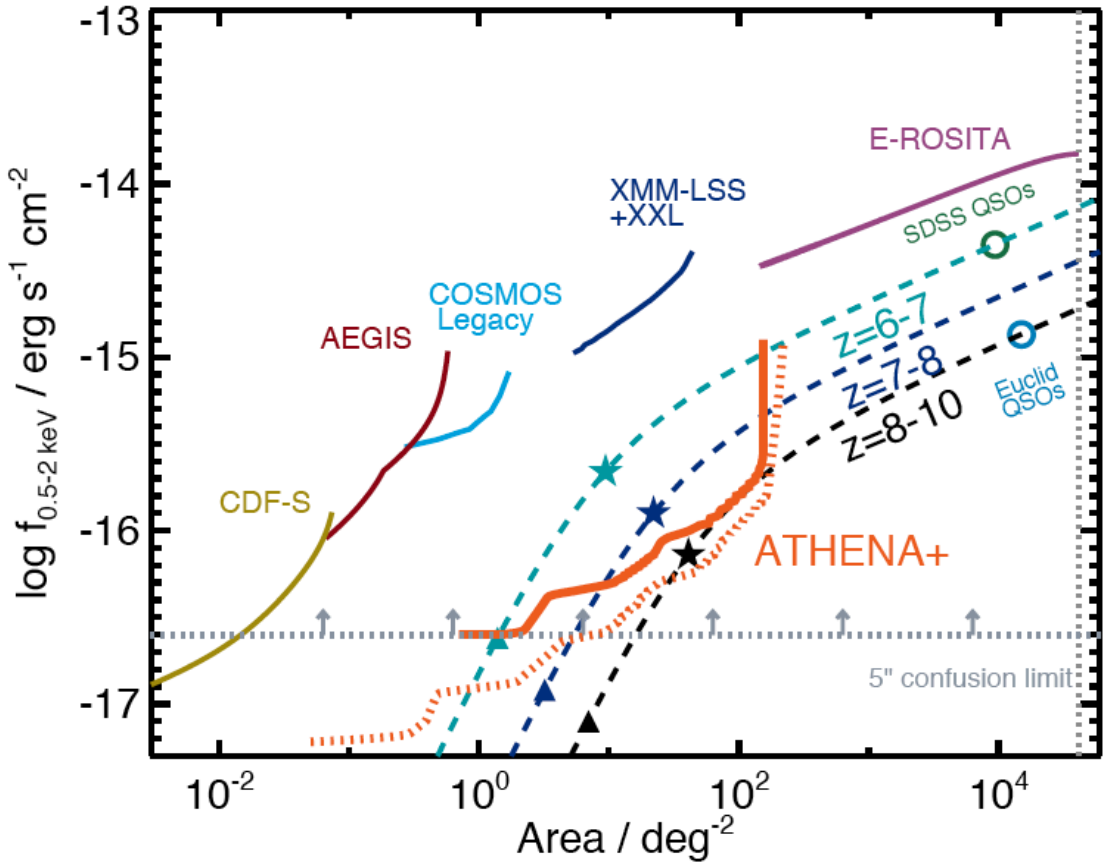

**Figure 2: Area/flux coverage achieved by *Athena+* compared to current surveys.** The dashed green, dashed blue and dashed black lines show the area/flux coverage that is required to detect >10 sources above a given limiting flux in the indicated redshift ranges. Circles indicate the approximate regimes probed by the SDSS and (forthcoming) Euclid optical/near-infrared spectroscopic surveys that identify luminous QSOs. The triangle and stars indicate where the flux limit corresponds to $L_X=10^{43}$ erg s$^{-1}$ and $L_X=10^{44}$ erg s$^{-1}$ at the given redshift – probing this regime is key to unveil the dominant, lower luminosity AGN population. With the baseline specification (5" PSF) and the multi-tiered 1-year survey strategy described in Section 3 (solid red line), *Athena+* will probe this new discovery space (that would require ~100 years of *Chandra* time to access) and revolutionise our knowledge of SMBH formation and growth in the early Universe. With the goal specification (3" PSF, dotted red), *Athena+* will probe to even greater depths, beyond that achievable with the deepest *Chandra* surveys to-date (CDF-S) and covering substantially larger areas.

The red curve in Figure 2 shows the area coverage to a given flux limit that could be achieved by *Athena+* with a multi-tiered survey strategy (4×1Ms, 20×300ks, 75×100ks, 250×30ks)[3] for a total observing time of 25Ms (i.e. 1 year, assuming an 80% observing efficiency). *Athena+* will probe a substantially new discovery space, placing strong constraints on the evolution of the XLF out to $z$~10 (see Figure 3). With the baseline *Athena+* specification, we predict we will identify over 400 $z$=6–8 AGNs, reaching down to luminosities of $L_X$=10$^{43}$ erg s$^{-1}$, and around 30 at $z$>8. However, as there are currently no observational constraints at $z$>6, the actual yields are uncertain – *Athena+* is required to constrain them. For more optimistic model predictions the yields could increase by an order of magnitude. In the most pessimistic extrapolation from the present XLF (e.g. assuming an exponential decay at all luminosities at $z > 3$) the number of detected AGN at $z > 6$ would be about a factor 2 lower (see orange dot-dashed line in Figure 3). Lower yields could rule out PopIII seeding mechanisms, which require high levels of accretion growth in the early Universe, and instead infer massive seeds growing at lower accretion rates (see Figure 1). Detection of AGNs out to $z \sim 10$ will place direct constraints on the minimum mass of SMBH seeds. For example, an AGN with $L_X \sim 4 \times 10^{43}$ erg s$^{-1}$ at $z \sim 10$ needs to be powered by a SMBH with mass $>10^7$ M$_\odot$, when the age of the Universe was only 0.48 Gyr. To reach this mass via Eddington-limited growth requires a seed mass $>10^5$ M$_\odot$. Thus, detection of such objects will help pin down the masses, growth rates and duty cycles of seed SMBHs. Measurements of the XLF at $z>6$ with *Athena+* will also place essential constraints on the level of ionising UV and X-ray emission from AGNs in the early Universe, thus constraining their contribution to reionisation.

---

[3] This strategy is also designed to identify large numbers of heavily obscured AGN at $z$=1–4, will yield enormous numbers of more typical, Compton-thin AGNs (~600,000) over all redshifts, and will provide an essential resource to search for high-$z$ X-ray clusters and groups (see Georgakakis, Carrera et al. 2013, Pointecouteau, Reiprich et al. 2013, *Athena+* supporting papers). We note that the deepest layer of our survey (4×1Ms) will be confusion limited with the baseline *Athena+* specification (5" PSF). This ultra-deep component is driven by the need for high quality X-ray spectral constraints to identify low-luminosity Compton-thick AGNs at $z$=1–4. Nevertheless, with the goal specification (3" PSF) this layer will reach beyond the depths of the current deepest *Chandra* fields and help place constraints on the very high-$z$ AGN population at even lower luminosity.

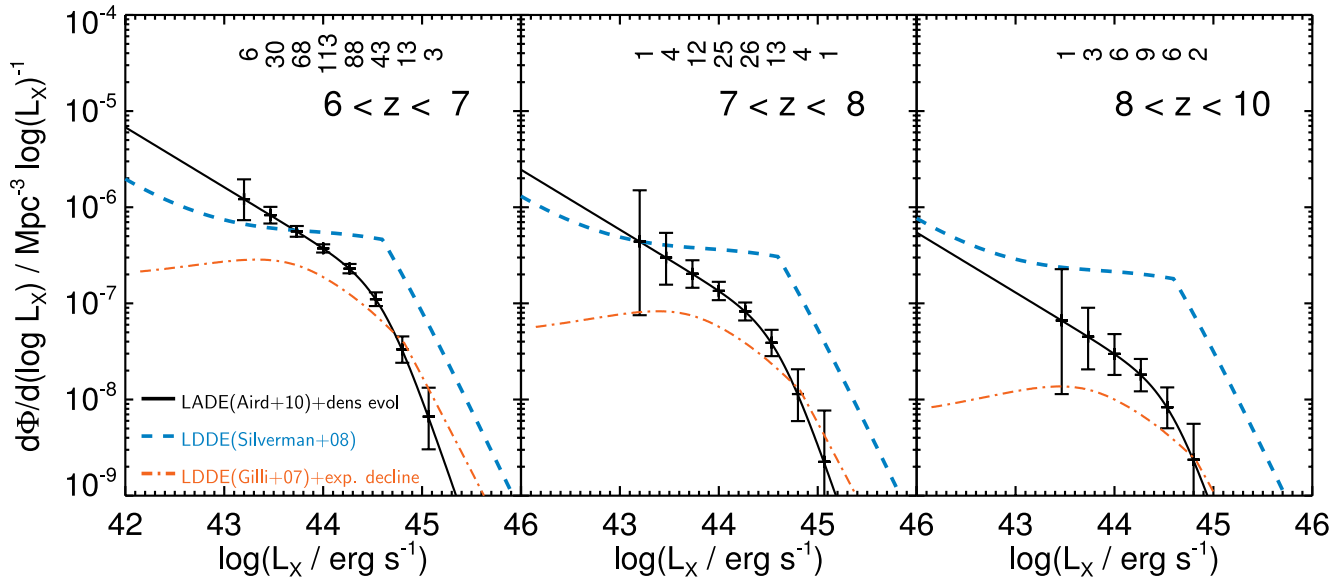

**Figure 3: Predicted measurements of the X-ray luminosity function (XLF) of AGNs at high-z with the 1-year *Athena+* WFI survey.** Points show the predicted binned measurements based on extrapolations of the Luminosity And Density Evolution (LADE) phenomenological model of Aird et al. (2010) with additional density evolution above z=3, which is shown by the black line. Error bars correspond to the Poisson uncertainties and the predicted number of sources in each bin is given at the top of each panel (assuming the baseline *Athena+* specification). We also show the extrapolations of luminosity-dependent density evolution models (LDDE) from Gilli, Comastri & Hasinger (2007, orange), including an exponential decline in the number density of AGNs at high redshifts, and Silverman et al. (2008, blue dashed). The actual shape of the XLF at these redshifts is unknown. *Athena+* will provide the first accurate measurements of the shape and evolution of the XLF out to $z \sim 10$, revealing the physical processes that drive SMBH growth at this early epoch and placing vital constraints on models for the initial formation and growth of SMBHs.

## 5. SYNERGIES WITH OTHER FACILITIES

The *Athena+* survey strategy has been designed to detect many hundreds of $z>6$ AGNs, without suffering from obscuration biases. However, a fully multi-wavelength approach, combining world class ground-based facilities and space missions across the electromagnetic spectrum, is required to understand the physical conditions in their host galaxies and their early evolution: it will make it possible to simultaneously measure, for the first time ever, host galaxy properties (e.g. stellar and gas masses and star formation rates) along with nuclear properties (e.g. accretion luminosity and SMBH masses) of the first lights in the Universe. This information is vital for understanding how SMBHs form, what fuels their subsequent growth, and what impact they have on the evolution of the first galaxies.

The proposed *Athena+* 1-year survey will play a fundamental role to enhance the scientific return of future wide field area surveys at longer wavelengths, which will predominantly probe the stars and dust in galaxies (Figure 4). Indeed, uniform, deep X-ray coverage over tens to hundreds of square degrees will be essential to pinpoint the most active

SMBH and serve as a starting point for deep and ultra-deep follow up. The key actors in the optical domain will be the Subaru Hyper-Suprimecam (HSC) and the LSST surveys, among others. By covering thousands of square degrees down to AB magnitudes of the order of 26 – 27 in the *u, g, r, i, z,* and *y* bands, these surveys will set the state of the art of optical imaging over large field area and identify the counterparts to luminous X-ray AGNs ($L_X > 10^{44}$ erg s$^{-1}$) at $z\sim6$–7 that will be detected by the shallow *Athena+* WFI survey (250×30ks). At higher redshift, the optical light will be completely absorbed by the intergalactic medium (IGM) and only large area near-infrared imaging such those provided by the Euclid-Wide survey (reaching near-infrared AB magnitudes of the order of 24 over 14,000 square degrees) has the capabilities to pinpoint luminous sources out to $z\sim10$. However, a large fraction of the numerous, lower luminosity, obscured $z>6$ AGNs identified in the medium-deep *Athena+* surveys are expected to have optical/near-infrared counterparts well below the above mentioned sensitivities, either because of redshift and IGM effects, and/or because of their expected lower bolometric luminosities. These *Athena+* surveys are well-matched to the areas and depths of the LSST-Deep and Euclid-Deep fields (reaching AB magnitudes ~28 and 26 in the optical and near-infrared, respectively, over ~40 square degrees), which will reveal the host galaxies of over 90% of the *Athena+* $z>6$ X-ray AGNs. Superior imaging capabilities (near-IR AB magnitude ~30) will also be offered by the James Webb Space Telescope (JWST) to reveal the counterparts of the AGNs in the deepest ~2 square degrees of the *Athena+* survey.

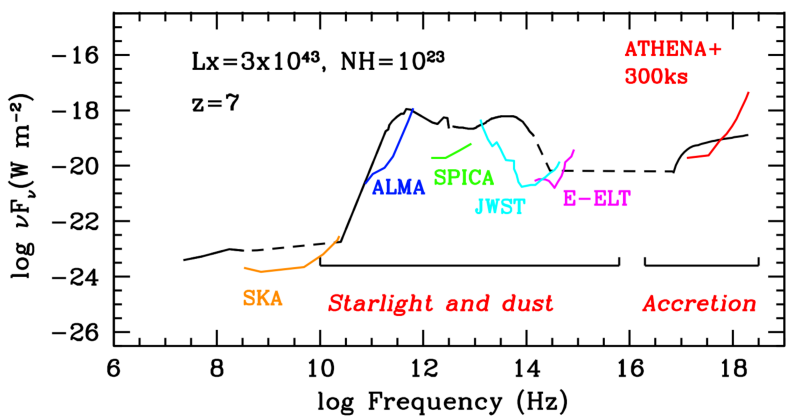

Figure 4: **Broad-band spectral energy distribution (SEDs) of a typical sources that can be observed with *Athena+* and other future multiwavelength facilities.** The average SED of an X-ray selected obscured ($N_H = 10^{23}$ cm$^{-2}$), low luminosity ($L_X \sim 3 \times 10^{43}$ erg s$^{-1}$) AGN (adapted from Lusso et al. 2011) redshifted to $z=7$ is shown as a solid line. In terms of bolometric luminosity, this object represents the typical yield of *Athena+* medium-deep surveys (e.g. 300 ks exposures, red line). The 3σ sensitivities (for a ~40ks exposure) of SKA, ALMA, SPICA, JWST and E-ELT are also shown, as labelled.

Further information on the hosts of the obscured $z>6$ AGNs found by *Athena+*, as well as accurate redshift measurements, will be provided by optical and infrared spectroscopic follow-up. While the JWST spectroscopic capabilities are limited by the relatively low collecting area, the planned European Extremely Large Telescope (E-ELT) instrumentation will deliver superb spectroscopic capabilities (down to $H_{AB} \sim 29$) and nicely complements the JWST imaging surveys. The cold dust component and the molecular gas dynamics at high redshifts will be probed by deep ALMA observations, using the [CII] 158 μm line and far infrared continuum shape. Furthermore, SKA is likely to detect a high fraction of the very high-*z* AGNs in the radio and will be vital for studying the physics of the accretion and emission processes for these early SMBHs.

*Athena+* may also hold the key to revealing the presence of heavily obscured, accreting SMBHs within samples of high-*z* galaxies that remain beyond the spectroscopic capabilities of E-ELT. A deep 1 Ms observation with the X-ray Integral Field Unit (XIFU) would provide ultra-deep, high-resolution X-ray spectroscopy and could unambiguously reveal and measure the redshift of moderate luminosity Compton thick AGNs at $z>8$ if a strong 6.4 keV (rest-frame) Fe K emission line is detected (see Figure 5). Detection of such a line would place constraints on the metallicity and could thus constrain the star-formation history of the host galaxy.

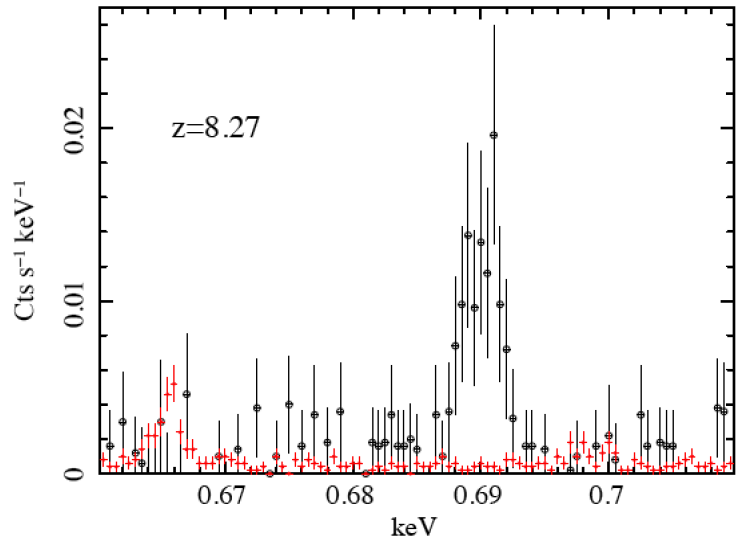

**Figure 5**: A simulated XIFU spectrum of a high-z (z=8.27) highly obscured, Compton thick AGN. The source is emitting a cold Fe K line at 6.4 keV with a ~0.7 keV equivalent width in the rest frame. The intrinsic source luminosity has been matched to that of NGC 6240 a prototype of a heavily obscured AGN in the local universe. The spectrum from a deep 1 Ms XIFU pointing is plotted in black along with the expected background spectrum in red. The iron line is observed as a highly significant (> 7σ) excess at ~0.69 keV. The original XIFU simulation is rebinned for plotting purposes.

## 6. REFERENCES

Aird J. et al., 2010, MNRAS, 401, 2531
Brusa M. et al., 2009, ApJ, 693, 8
Civano F. et al., 2011, ApJ, 741, 91
Di Matteo T. et al., 2012, ApJ, 745, L29
Dubois Y. et al., 2013, MNRAS, 428, 2885
Dunkley J. et al., 2009, ApJS, 180, 306
Fan X. et al., 2003, AJ, 125, 1649
Faucher- Gigue`re C. et al., 2008, ApJ, 688, 85
Fiore F. et al., 2012, A&A, 537, A16
Georgakakis A., Carrera F., et al., 2013, *Athena+* supporting paper, Understanding the build-up of supermassive black holes and galaxies at the heyday of the Universe, available from http://www.the-athena-x-ray-observatory.eu/
Gilli R., Comastri A. & Hasinger G., 2007, A&A, 463, 79
Li Y. et al., 2007, ApJ, 665, 187
Lusso E. et al., 2011, A&A, 534, A110
Meiksin A., 2005, MNRAS, 356, 596
Mortlock D.J. et al., 2011, Nature, 474, 616
Pointecouteau E., Reiprich T. et al., 2013, *Athena+* supporting paper, The Evolution of Galaxy Groups and Clusters, available from http://www.the-athena-x-ray-observatory.eu/
Riechers D.A. et al., 2013, Nature, 496, 329
Schawinski K. et al., 2011, ApJ, 743, L37
Silverman J.D. et al., 2008, ApJ, 679, 118
Trichas M. et al., 2012, ApJS, 200, 17
Ueda Y. et al., 2003, ApJ, 598, 886
Volonteri M. & Begelman M.C., 2010, MNRAS, 409, 1022
Volonteri M. & Gnedin N.Y., 2009, ApJ, 703, 2113
Xue Y.Q. et al., 2011, ApJS, 195, 10